\documentclass[pdflatex,sn-mathphys-num,a5paper]{sn-jnl}
\usepackage{graphicx}%
\usepackage{multirow}%
\usepackage{amsmath,amssymb,amsfonts}%
\usepackage{amsthm}%
\usepackage{mathrsfs}%
\usepackage[title]{appendix}%
\usepackage{xcolor}%
\usepackage{textcomp}%
\usepackage{manyfoot}%
\usepackage{booktabs}%
\usepackage{algorithm}%
\usepackage{algorithmicx}%
\usepackage{algpseudocode}%
\usepackage{listings}%
\usepackage{subcaption}
\usepackage{setspace}
\usepackage{gensymb}
\usepackage{lineno}
\graphicspath{{./Figures/}}
\DeclareGraphicsExtensions{.pdf,.tiff,.png}
\begin{document}

\title[Article Title]{Spatially-Localized Second Harmonic Generation via Spin Wave Concentration in Patterned YIG Structures}

\author[1]{\fnm{Stephanie R.} \sur{Lake}}

\author[1]{\fnm{Marc} \sur{Eger}}

\author[1]{\fnm{Philipp} \sur{Geyer}}

\author[1]{\fnm{Rouven} \sur{Dreyer}}

\author[1]{\fnm{Seth W.} \sur{Kurfman}}

\author*[1,2,3]{\fnm{Georg} \sur{Schmidt}}\email{georg.schmidt@physik.uni-halle.de}

\affil[1]{\orgdiv{Institut für Physik}, \orgname{Martin-Luther-Universität Halle-Wittenberg}, \orgaddress{\city{Halle}, \postcode{06120}, \country{Germany}}}
\affil[2]{\orgdiv{Interdisziplinäres Zentrum für Materialwissenschaften}, \orgname{Martin-Luther-Universität Halle-Wittenberg}, \orgaddress{\city{Halle}, \postcode{06120}, \country{Germany}}}
\affil[3]{\orgdiv{Halle-Berlin-Regensburg Cluster of Excellence CCE} \country{Germany}}


\abstract{\textbf{
The anisotropic dispersion and inherent nonlinearity of magnetostatic spin waves in thin films and confined structures provide unique opportunities for implementation in next-generation magnonic devices for data and signal processing. A particular challenge is to establish an effective means to locally generate higher harmonics and subsequently exploit them while avoiding extraneous nonlinear losses.
Here we demonstrate that deterministically and locally tuning the dispersion relation by geometric confinement through standard patterning processes, allows the creation spatially localized, high-intensity magnons hundreds of µm or even further from the excitation source. The local intensity obtained in passive, lithographically patterned YIG funnel structures is sufficient to achieve second harmonic generation in localized regions via conventional magnon scattering processes. We verify these effects are truly nonlinear processes by direct measurement and comparison of the 1-$\omega$ and 2-$\omega$ magnon signals as determined by highly sensitive frequency- and spatially-resolved SNS-MOKE technique. This lays the foundation for using similar devices in future magnon-based infrastructures to localize and enhance sensitivity of readout, downstream magnon-based logic operations, and for other higher harmonic generation-related phenomena and low-power magnonics applications.
}}


\maketitle

\section{Introduction}\label{sec1}
The magnetic excitations of magnetically-ordered spin systems, known as spin waves (or their quanta, magnons), enable nearly dissipationless wave-based operations at low driving power and intermediate transduction of information between systems with different carriers of information (e.g. spin, charge, vibration, etc...)~\cite{Chumak2022, Flebus2024, Engelhardt2022}. Moreover, the high (radio) frequency dynamics of spin waves, as well as their intrinsic nonlinear nature, provide exciting directions for integration and exploitation of magnons in microwave devices~\cite{Ustinov2019}. 
The power threshold of non-linear spin wave dynamics, which are described by the Landau-Lifshitz-Gilbert equation, typically requires large magnon population densities.  This traditionally stimulates most nonlinear mechanisms to occur nearby the magnon excitation source (i.e. waveguide/antenna), thereby decreasing the efficiency as the distance away from the source increases. This limitation makes separating nonlinear excitations from the source and restricting them to isolated regions a technological challenge\cite{Hula-APL2020, Divinskiy-NC2019, Bauer2015}.

In principle, this challenge can be addressed by a device that is capable of collecting low-intensity, spread-out spin waves into a localized region where they exceed the threshold for nonlinearities while simultaneously minimizing propagation and interaction losses. For this purpose, one can draw inspiration from light-collecting optics devices that rely on refraction (in lenses), reflection (in mirrors), or diffraction (for example in Fresnel zone plates) to control the propagation direction of light. 
In principle, all of these effects can be realized for magnon-based optics~\cite{Kiechle2022, Whitehead2019, Graefe2020, Vogel2020}; however, the anisotropic spin wave dispersion constrains the changes of the direction of the group velocity that are necessary for efficient focusing. Thus, this anisotropic behavior must be carefully considered if one wants to  tune the spin wave's propagation direction. This work focuses on refractive principles to control the propagation direction of plane spin waves in order to concentrate their intensity in a small spot. 

Fundamentally, the magnon dispersion within a magnetic material is explicitly dependent on the geometry and the local (effective) magnetic fields~\cite{Stigloher2016}. Accordingly, these two parameters can be used to control the propagation direction and achieve the necessary concentration.

A review of refractive spin wave optics explorations shows that a common approach is to locally modify the magnetic field, and thus the spin wave dispersion, in order to bend the path of a magnon~\cite{Davies2015}. Numerous examples of such approaches have been reported, including external fields with micro-scale coils \cite{Bok2024, Moore2015}, adding ferromagnetic elements \cite{Jorzick2002, Haldar2016, Qin2021, Dreyer2021}, and patterning artificial structures such as antidots and gratings \cite{Gieniusz2013, Gieniusz2017, Graefe2020}.  Common to all these devices is their rather limited  increase in intensity, which is partially due to the non-optimized focusing effects in the utilized structures but also because these devices all rely on metallic ferromagnets in which the spin wave damping is large~\cite{Toedt2016,Tanabe2014}. As a result, any enhancement of the spin wave intensity due to focusing/concentration is mostly canceled out by the losses from damping that limits propagation lengths. 

These limitations can be overcome through the use of low-loss magnonic materials, such as yttrium iron garnet (YIG). Nevertheless, the challenge remains to deterministically modify the spin wave dispersion and optimize the device geometry to restrict non-linear behavior to a selected area. Established patterning techniques \cite{Hauser2016, Heyroth2019} enable the nano-fabrication of YIG-based devices that directly satisfy both of these demands.

\begin{figure}
    \centering
    \includegraphics[width=\textwidth]{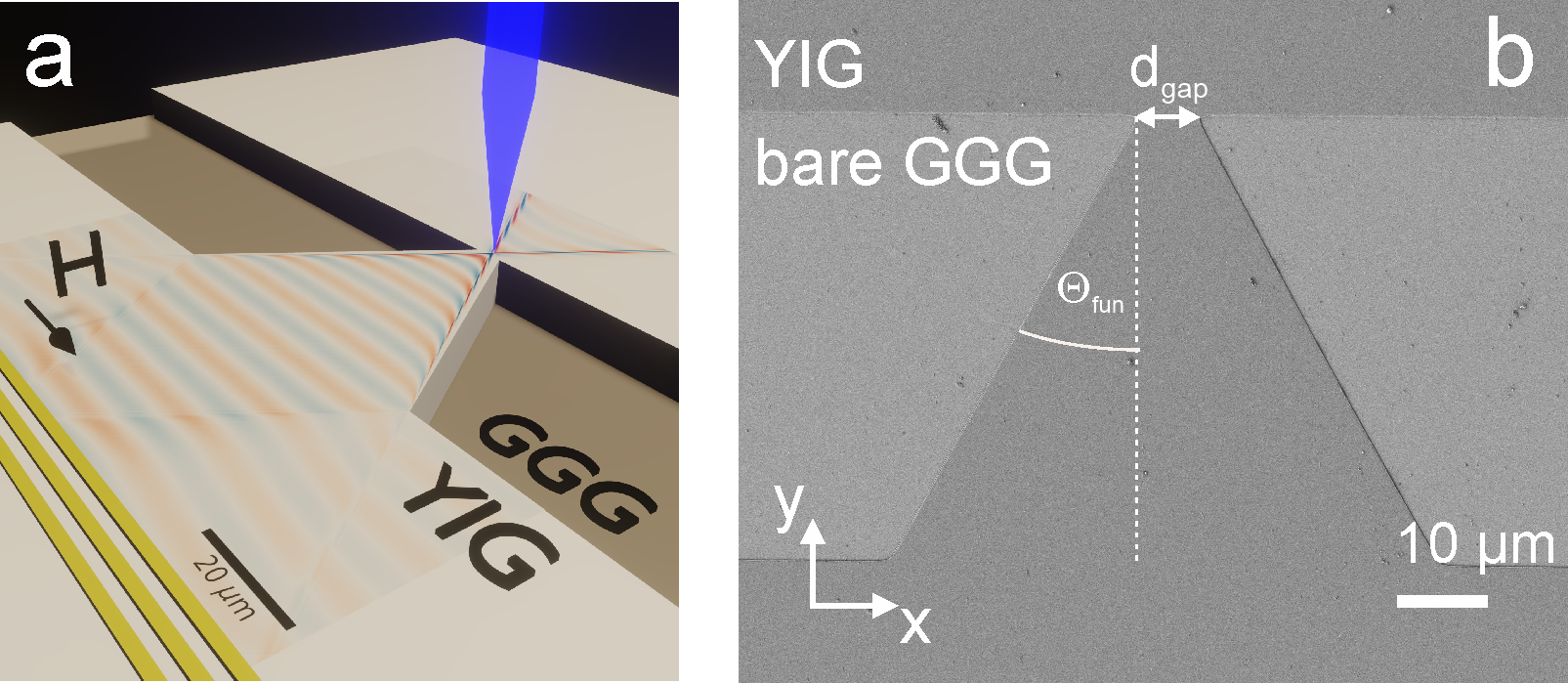}
    \caption{(a) shows a magnon concentrator with a simulated spin wave pattern detected by SNS-MOKE microscopy. (b) Scanning electron microscopy image of a YIG-based magnon concentrator, with labels for the various parameters defining the device.}
    \label{fig:INTRO}
\end{figure}

Here, we showcase a device comprised of lithographically-patterned YIG in the shape of a funnel, which we coin as a \textit{magnon concentrator}, capable of directing a plane propagating spin wave wavefront to a focal point. It should be noted while focusing is essential, imaging capability is not necessary, making optical non-imaging concentrators \cite{Madala2017} rather than lenses the corresponding device from light optics. As shown later, the device enables us to create a focus much smaller than the nominal diffraction limit would allow. The realized design, as seen in Fig. \ref{fig:INTRO}, takes advantage of the modified anisotropic spin wave dispersion in the presence of a spatially-varying effective magnetic field near a boundary in order to redirect and concentrate spin waves. The material choice and device design allows for the spatial confinement and enhancement of spin wave intensity that exceeds the losses due to propagation-related decay mechanisms, and as a result allows to demonstrate an enhancement of the spin wave amplitude by a factor of $~23$. Moreover, for a range of external bias fields and driving frequencies this enhancement of spin wave intensity in a localized region allows coherent second-harmonic generation (SHG). This SHG of propagating spin waves is directly confirmed via phase-resolved super-Nyquist sampling magneto-optical Kerr effect (SNS-MOKE) microscopy in regions far from the excitation source ($\sim 100$ µm). Our findings are supported by micromagnetic simulations, and we provide an accessible model that explains the spin wave steering near the magnetic material boundaries. All together these results show that the spin wave intensity can be passively increased by several orders of magnitude, thereby directly enabling highly sensitive readout of low intensity magnons. Furthermore, such device design principles and applications of localized non-linear frequency conversion may enable other higher-harmonic generation (HHG) applications including, for example, signal demultiplexing at RF device outputs.

\section{Theory of Magnon Concentration}\label{sec2}

In order to understand the necessary physics for the realization of a magnon concentrator that controls the magnon group velocity, one must first consider the anisotropic dispersion relation for spin waves. The dispersion for a magnetic thin film homogeneously magnetized in the film plane is well described by the formalism established by Kalinikos and Slavin~\cite{Kalinikos-SPJ1981} and accurately predicts the dipole-exchange magnons in the magnetic film. In the typical Damon-Eschbach (DE) geometry, the magnon wavevector is perpendicular to the magnetic field ($\vec{k}\perp\vec{H}$), the field density ($\vec{k}\perp\vec{B}$), and the magnetization ($\vec{k}\perp\vec{M}$). These magnons can additionally quantize along the film normal, leading to a higher frequency for the same $k$-vector. 
~\cite{Kalinikos1986}. In our device, these quantized modes are vital for the SHG. The dispersion relation within a ferromagnetic film with unpinned boundary conditions is:
\begin{equation} \label{eq:omega_fn}
    \omega = \gamma \mu_0 \sqrt{(H_{\mathrm{eff}}+A M_{\mathrm{sat}} k^2)(H_{\mathrm{eff}} + A M_{\mathrm{sat}} k^2 + M_{\mathrm{sat}} F )}
\end{equation}
where $\omega/2\pi=f$ is the spin wave precession frequency, $\gamma/2\pi \sim 28$ GHz/T is the gyromagnetic ratio, and $\mu_0$ is the vacuum permeability. Furthermore, it is determined by the effective magnetic field $\mu_0 H_{\mathrm{eff}}$, saturation magnetization $\mu_0 M_{\mathrm{sat}}$, exchange constant $A$, wavenumber $k$, and anisotropic function $F$. This function $F= F(k,M_{eff}, H_{eff}, \phi)$ is related to $\phi$, the in-plane angle between $\vec{H}_{eff}$ and the in-plane component of the $k$-vector by:
\begin{equation} \label{eq:F_fn}
    F = 1 + P (1 - P)\left[ \frac{M_{\mathrm{sat}}}{H_{\mathrm{eff}}+A M_{\mathrm{sat}} k^2} \right]\sin^2 (\phi)-P \cos^2(\phi)
\end{equation}
Notably in Eq.~\ref{eq:F_fn}, if $H_{\mathrm{eff}}$ or $M_{\mathrm{sat}}$  changes, then to conserve frequency, the incoming $\vec{k}$ must change in-plane direction, length, or both. 

\begin{figure}
    \centering
    \includegraphics[width=\linewidth]{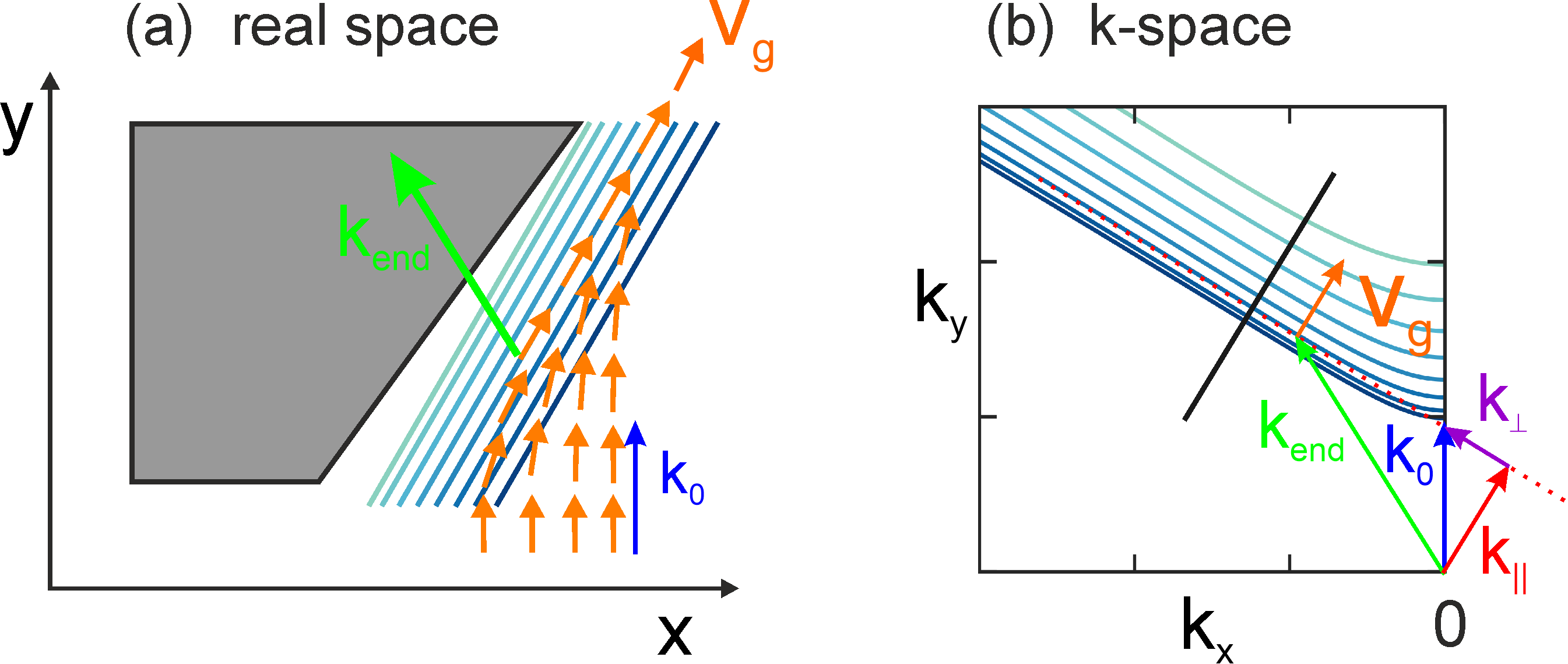}
    \caption{Change of $k$-vector and group velocity for DE modes in the device. In real space the orange arrows show the group velocity changing along the magnon path. In $k$-space the orange arrow only indicates the final group velocity which is parallel to the iso-field line (black line). 
    $k_0$ (blue) is the initial $k$-vector in y-direction. $k_{end}$ (green) is elongated and tilted towards the boundary. The blue lines in (a) are iso-field lines in real space while in (b) they represent iso-frequency lines for constant respective magnetic field in $k$-space. When the original spin wave with $k$ and $v_g$ in x-direction approaches the boundary, $k_{||}$ (red) is conserved while $k_\perp$ changes and grows until an iso-frequency line is reached where the group velocity is parallel to the iso-field line. In real space, a broad magnon beam is thus concentrated on the corresponding iso-field line where  $v_g$ is along this line.
    }
    \label{fig:THEORY}
\end{figure}

To visualize the possible $k$-vectors for a certain frequency, one can plot so-called iso-frequency lines (Fig. \ref{fig:THEORY}). Each line shows the different $k$-values and their respective directions in space that are allowed for a single given frequency. The different lines belong to different respective magnetic fields. Each point on one of these lines represents a $k$-vector (starting at the origin) and a group velocity which is perpendicular to the iso-frequency line at this point. Note that along an iso-frequency line, $\omega$ is always constant so the tangential component of $\partial \omega/\partial k$ must be zero. As a consequence, the group velocity $\vec{v}_g = \vec{\nabla}_{\vec{k}}(\omega)$ - if non-zero - can only be perpendicular to the iso-frequency line.
We now consider a $k$-vector ($\vec{k}_0$, blue) in the $y$-direction, whose group velocity $\vec{v}_g$ is initially parallel to $\vec{k}_0$. We also add to our system a straight lateral boundary to air.

This boundary must be aligned to satisfy certain conditions. At the boundary, a demagnetizing field occurs. Because of this demagnetizing field, a gradient of the magnetic field is observed close to the boundary with the magnetic field decreasing towards the boundary. This gradient is almost but not completely perpendicular to the boundary itself. As a result we have lines of constant magnetic field (iso-field lines, different shades of blue) which are almost (but not completely) parallel to the boundary. The functionality of our device requires that these iso-field lines are perpendicular to the iso-frequency lines in the area where those are more or less straight.
Along these iso-field lines we have translational symmetry.

This translational symmetry imposes the conservation of the $k$-component parallel to the iso-field lines ($k_{||}$, Fig.\ref{fig:THEORY}b, red arrow) when the spin wave approaches the interface. At the same time, the magnons that are approaching the interface with $\vec{v}_g$ and $\vec{k}$ must change both these properties when the magnetic field decreases. Because $k_{||}$ is conserved, only the $k$-component perpendicular to the iso-field lines ($k_{\perp}$, purple) can change, so $\vec{k}$ moves on the dotted line by increasing in length and rotating towards the iso-field line.
This process continues until the dotted line lies on top of, and thus parallel to, the current iso-frequency line. At this point, the group velocity $\vec{v}_g$ (orange) is parallel to the respective iso-field line which is perpendicular to the iso-frequency line. As a consequence, the magnetic field that the magnon experiences while propagating, no longer changes and both $\vec{v}_g$ and $\vec{k}$ remain constant. From the diagram, one can also infer that this rotation of $\vec{k}$ toward the iso-field lines increases for smaller $|{\vec{k}_0}|$. Theoretically for $|{\vec{k}_0}|\rightarrow 0$ this final value of $\vec{k}$ is perpendicular to the iso-field lines (parallel to the iso-frequency line). 
In Fig. S1 (supplementary material) scenarios for different angles are shown together with micromagnetic simulations demonstrating that stronger deviations from the ideal case lead to reflections for which the conservation of $k_{||}$ and the change of $k_{\perp}$ can lead to interesting changes in $\vec{k}$ and $\vec{v}_g$. 

These considerations show us that because there is only one set of iso-frequency lines for one frequency, each frequency determines an angle of a boundary that allows us to concentrate all magnons of this frequency initially traveling in the positive $y$-direction on a single line of constant magnetic field and thus create a very narrow high intensity magnon beam from a broad low intensity wave front. The effect is mirror symmetric and mirroring the boundary with respect to the $y$-axis creates a funnel with a second mirrored magnon beam. The two beams cross at a focal point where they automatically interfere constructively thereby doubling the amplitude. Extending the length of the boundary allows to collect more magnons and the final intensity is only limited by the propagation losses that for a large device will finally surpass the collection efficiency.
At first glance the theory suggests there is exactly one ideal boundary, making the device extremely efficient under perfect conditions but extremely sensitive to small geometric deviations. In reality, the condition for the boundary angle is somehow softened because the iso-frequency lines can neither be perfectly straight nor perfectly parallel. On the one hand this means that for one frequency and one angle not all values of $|{\vec{k}_0}|$ will yield the same efficiency, but at the same time a slight misalignment will not make the device useless but only shift the range of suitable $|{\vec{k}_0}|$. Finally the iso-field lines change direction at the funnel's exit. In this region, the magnon beam can undergo changes in $\vec{v}_g$ and $\vec{k}$ before again reaching an area with a new constant field and magnetization behind the funnel. Because the magnon beam is narrow and ideally has a uniform $\vec{k}$, it deflects and does not broaden.  This is for example visible in Fig. \ref{fig:EXAMPLE} where the exiting beams are not parallel to the funnel sides, however, match the $k$-vector for second harmonic generation.

Considering the above premise, optimizing the performance of the magnon concentrator is mainly a matter of geometry. As we show below, such optimization enables us to achieve an increase in magnon amplitude up to 23.4 times in the experiment (corresponding to an increase in intensity of $543$). 

\begin{figure}
    \centering
    \includegraphics[width=\linewidth]{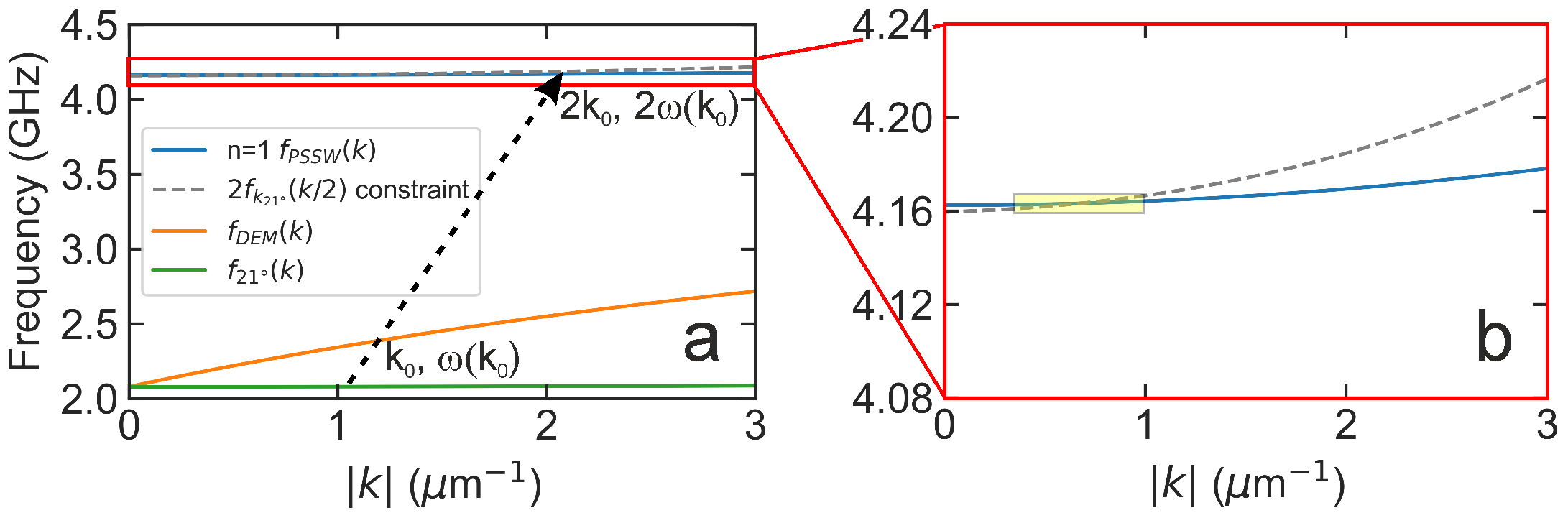}
    \caption{Plots of the analytical dispersion relations calculated from Eq. \ref{eq:omega_fn}. (a) shows the plot containing all the functions of interest. (b) is a zoomed-in plot showing both $2f_{k_{21\degree}}(k/2)$ and $f_{\text{PSSW}_{21\degree}}(k)$ for $n=1$. In the yellow shaded area the two functions are close enough to overlap within the frequency linewidth.}
    \label{fig:DISPERSIONS}
\end{figure}

This result indicates that it may be possible to locally increase the intensity above the threshold for non-linear magnonics, with second harmonic generation being the likeliest phenomenon. Besides high magnon intensity, SHG requires two conditions to be fulfilled: in the process both energy and momentum must be conserved. Energy conservation is implicit with the creation of a magnon with $2\omega$ out of two magnons with $\omega$. But also

\begin{equation}
    \vec{k}_{\omega}+\vec{k}_{\omega}=\vec{k}_{2\omega} 
    \label{SUM_k}
 \end{equation}
must be satisfied. This is only possible if the magnon dispersion exhibits suitable pairs of magnon energies and $k$-vectors, typically on different dispersion branches~\cite{Demidov2011}. This can be checked by using one dispersion branch $\omega(\vec{k})$ and plotting $2\omega(\vec{k}/2)$ for the same branch (see Fig. \ref{fig:DISPERSIONS}). A magnon at a $k$-vector of $2\vec{k}_0$ on this new curve with $\vec{k}_0$ being any $k$-vector on the original curve automatically has the frequency $2\omega_0$ with $\omega_0$ being the frequency for the vector $\vec{k}_0$ on the first curve as indicated by the black arrow. If this new curve crosses any other dispersion branch, the conditions mentioned above are automatically met and SHG is possible as has also been shown by Nikolaev et al.~\cite{Nikolaev2024}.

The high intensity magnons collected in our device fulfill these two conditions. The fact that $\vec{\nabla}_{\vec{k}}$ is almost perpendicular to $\vec{k}$ implies that $\partial{\omega}/\partial{|k|}\approx 0$ so the dispersion curve for the concentrated magnons is almost flat (Fig. ~\ref{fig:DISPERSIONS}). This also suggests that not only $2\omega(\vec{k}/2)$ is flat, but also the dispersion branch for the corresponding perpendicular standing spin wave.

By such a modification of the spin wave dispersion, a single crossing point of two branches with different group velocity transforms into a broad range of wavenumbers (still for a narrow range of $\omega$) for which these two branches are merely on top of each other. A finite frequency linewidth of the spin waves allows for SHG over an extended range of $k$-vectors, a feature which is unique to this magnon concentrator device. 

\section{Results \& Discussion}\label{sec3}
\subsection{Simulations}
To validate the conjectures of the model discussed above, micromagnetic simulations were carried out in order to find a suitable set of device parameters and experimental conditions to achieve highest magnon intensities and simultaneously fulfill the conditions necessary for SHG. These simulations used the micromagnetic simulation software $\mathsf{MuMax3}$ \cite{Vansteenkiste2011} and  standard parameters for YIG  (see Methods). 
The simulations show that indeed for a certain funnel angle the mechanism described above can lead to a large increase in magnon intensity. The concentration factor achieved in the simulations can be as high as 17.5 (Fig.\,\ref{fig:Max}  in amplitude corresponding to a 340-fold increase in intensity. Because each simulation run only allows for the test of one discrete set of parameters (frequency and magnetic field) this maximum value is not necessarily the maximum possible one which may appear between two tested parameter sets. Indeed as we show below, the experiment even yields a larger value.

\subsection{Experimental Verification}\label{subsec1}

\begin{figure}
    \centering
    \includegraphics[width=\linewidth]{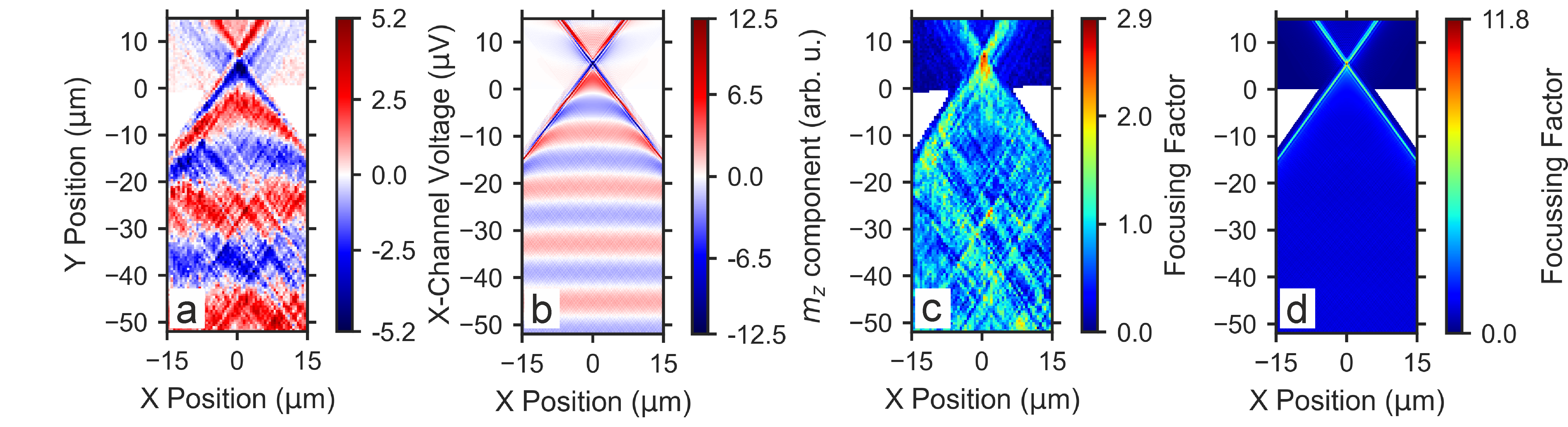}
    \caption{Simulation and experiment to demonstrate the operational principle of a funnel concentrator. The left hand figures show the oscillating amplitude of the dynamic $m_z$ magnetization component while the right hand figures show the magnitude of the spin wave (which is always positive) normalized to the magnitude at the entrance of the funnel. This value defines the focusing factor which is at its maximum in the focus. The maximum focusing factor in this experiment is ~2.9 and in theory is 10.7. The device parameters were not perfectly matched, resulting in different $k$-values for the magnons and a different focusing factor. Details of the measured and simulated device parameters can be found in the supplementary material.
	} 
    \label{fig:MuMax-Vs-MOKE}
\end{figure}

Following the general device parameters identified from the simulations, YIG-based magnon concentrators were fabricated using the techniques derived from \cite{Heyroth2019} (see Methods). Although different geometries are possible depending on the frequency range of operation, we only discuss three sets of parameters to demonstrate the functionality of the funnel concentrator.

The local measurement of the spin wave amplitude in our devices was measured using scanning super-Nyquist-sampling MOKE (SNS-MOKE) that allows for simultaneous detection of amplitude and phase information of the dynamic out-of-plane magnetization component at arbitrary frequencies~\cite{Dreyer2021}. Simultaneous measurements of linear and non-linear frequency components~\cite{Koerner2022} are enabled by using lock-in demodulation at twice the excitation frequency corresponding to second harmonics.

In Fig.~\ref{fig:MuMax-Vs-MOKE} a full Kerr map of the spatially-resolved spin wave amplitude for an example spin wave concentrator (sample A, $\Theta_{fun}=35^{\circ}, d_{gap}=12\,\mu m$, $L_{fun} = 50 \text{ $\mu$m}$, see supplement for more experimental details) is shown and compared to micromagnetic simulations. Figure~\ref{fig:MuMax-Vs-MOKE} shows that in good agreement with simulations, the propagating spin wave is concentrated in two high intensity beams parallel to the funnel boundaries leading to maximum intensity in the focus where these two beams cross. In the following we define the focusing factor as the maximum signal amplitude in the focus divided by the averaged amplitude in a reference area at the point where spin waves enter the funnel. For the example device this ratio is 2.9 in the experiment. The highest efficiency of all measured devices, however, was measured in a different device (sample B, $\Theta_{fun}=33^{\circ}, d_{gap}=6\,\mu m$,$L_{fun} = 50 \text{ $\mu$m}$, see supplement for more experimental details) shown in Fig.~\ref{fig:Max}. To limit measurement time for the time consuming Kerr maps, in most of the cases the experiment did not map the whole device area but only the areas around the point of concentration and around the reference point at the beginning of the funnel to determine the focusing factor. For the device of maximum efficiency (Fig.~\ref{fig:Max}) a focusing factor of 23.4 was achieved. For the intensity which goes with the square of the amplitude this means an increase by a factor of approx. 547. It is interesting that the size of the focus seems to fall below the diffraction limit when compared to the incoming wave vector. Note, however, that during the concentration, the wave vector increases dramatically and so for the $k$-value in the focus the diffraction limit may still apply. For this device, the experimentally obtained focusing factor is even higher than that obtained in the simulation (17.5), again because small deviations of the parameters can lead to large changes in the focusing factor and a continuous variation of parameters is neither possible in the experiment nor in simulation.
\begin{figure}[t]
    \centering
    \includegraphics[width=\linewidth]{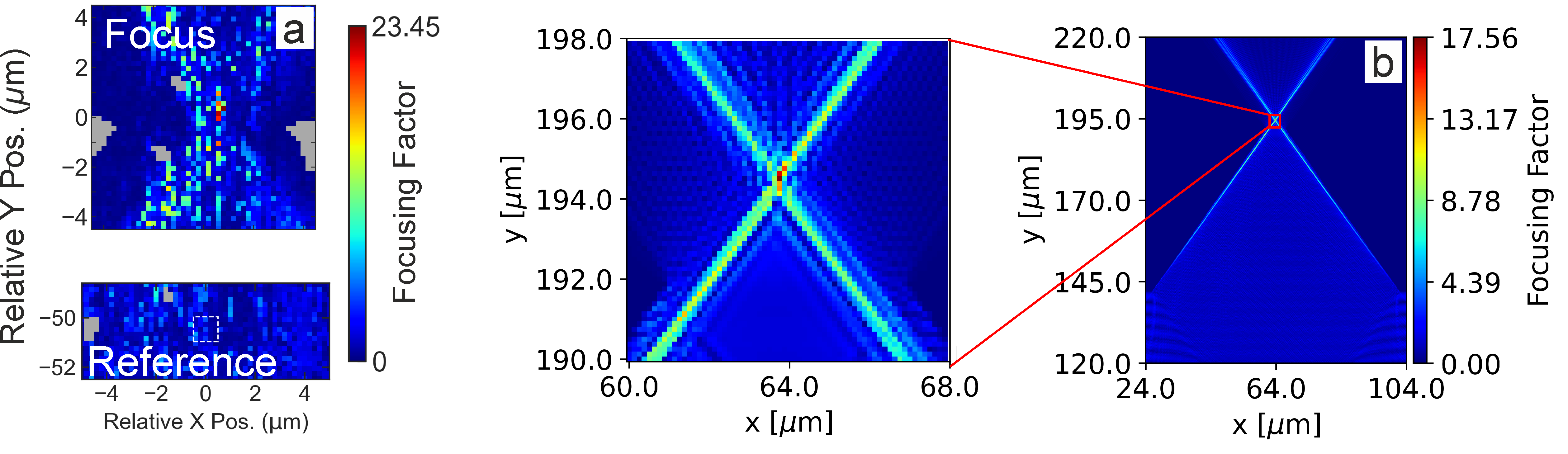}
    \caption{(a) Magnitude map for the device that showed the maximum focusing efficiency in the experiment and (b) corresponding simulation. The measurement shows only the focus area and reference area before the funnel. For the simulation, the focus area is also shown as a blow-up to demonstrate that the focus can be extremely small. The focusing efficiency in the measurement is even larger than in the simulation because it is extremely sensitive to the chosen parameters. The focusing factor is obtained by dividing the maximum values in the focus (which is basically noise free) by the averaged intensity in the reference area where noise must be reduced to obtain a reliable result.
	} 
    \label{fig:Max}
\end{figure}

\begin{figure}
    \centering
    \includegraphics[width=\linewidth]{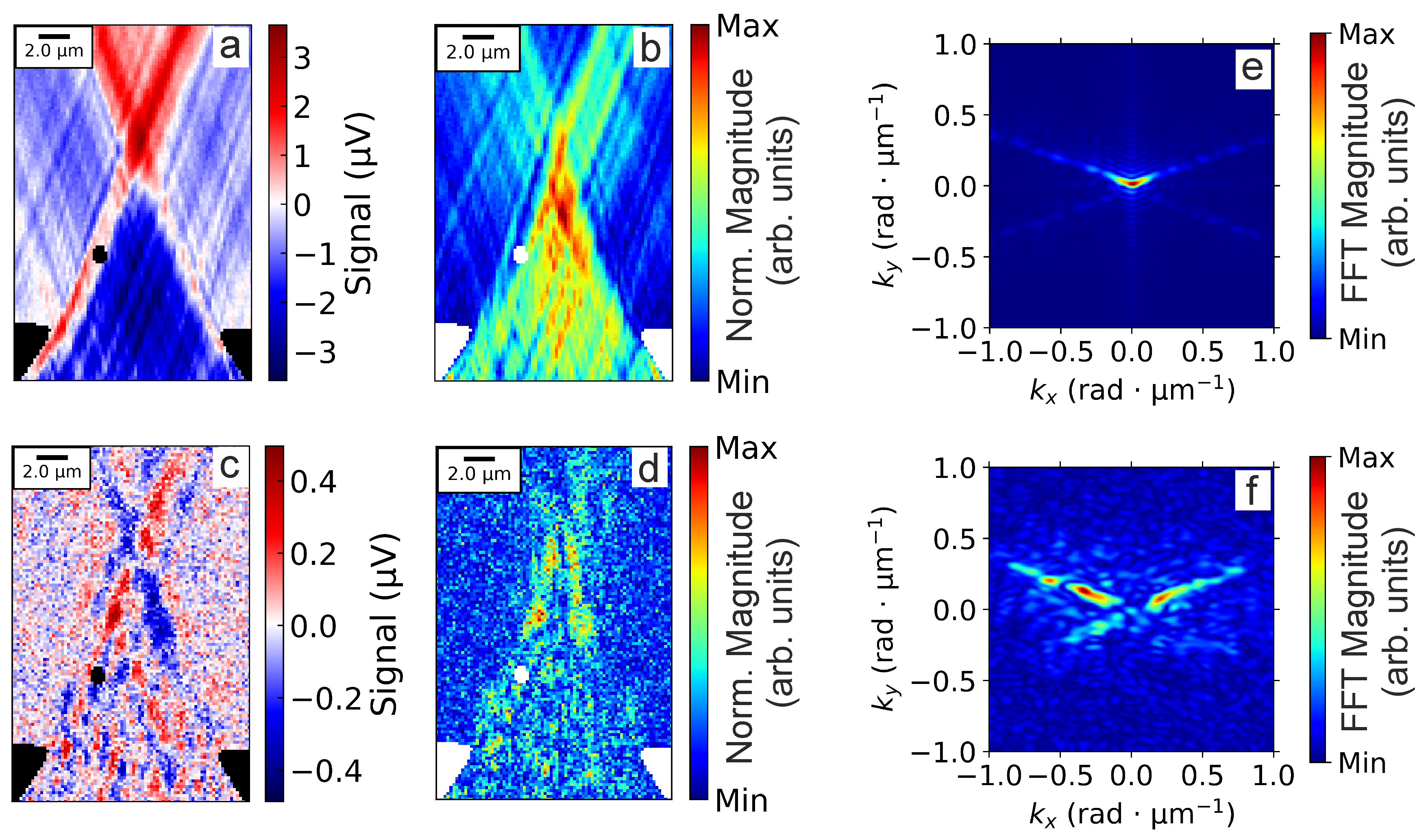}
    \caption{(a-f) Measurements demonstrating SHG in a funnel structure. (a) Phase resolved image of magnons and (b) magnon intensity in an optimized spin wave concentrator measured at $1f$ by SNS-MOKE. (c) Phase resolved image of magnons and (d) magnon intensity in an optimized spin wave concentrator measured simultaneously at $2f$. (e and f) show the respective FFT for the focus area. $\vec{k}$ points at $\pm 21\degree$ left and right. For $2f$, $\vec{k}$ is pointing in the same directions but the length is approx. doubled.}
    \label{fig:EXAMPLE}
\end{figure}

From the dispersion relation (Fig. \ref{fig:DISPERSIONS}) we can see that the way that the conditions for second harmonic generation are fulfilled also depends on the direction of the $k$-vector and thus ultimately on the funnel angle $\Theta_{fun}$. Indeed, we do not observe SHG in device B that showed the highest focusing factor but in a third device (sample C, $\Theta_{fun}=35^{\circ}, d_{gap}=10\,\mu m$,$L_{fun} = 50 \text{ $\mu$m}$, see supplement for more experimental details). Prior micromagnetic simulations have indicated SHG for such a device around a frequency of 2 GHz at a corresponding bias field of 26 mT. The experiment confirms the SHG under these experimental conditions, as depicted in Fig. \ref{fig:EXAMPLE}.  


In addition to the phase resolved imaging and the local spin wave intensities for the excitation frequency and the second harmonic, two more panels show the 2D FFT data from the measured images. The FFT exhibits a range of $k$-values at dominant angles ($159\degree$ and $21\degree$, respectively) rather than a single value. This effect results from the flat dispersion which yields a range of $k$-vectors for a single frequency within the finite line width combined with the fact that on the corresponding iso-frequency line, there is a range of $k$-vectors with parallel respective group velocity in the beam direction. The same is observed for the second harmonic, now with larger $k$-values but the same dominant angle as determined by Eq.~\ref{SUM_k}.

\subsection{Robustness}\label{subsec2}
As indicated before, the fact that we observe SHG over an extended magnetic field range and a larger range of $k$-vectors rather than a small one \cite{Demidov-PRB2011} is explained by the flat dispersion (Fig. \ref{fig:DISPERSIONS}). It is interesting to note that despite these results the frequency window for SHG is much smaller than for example in ~\cite{Nikolaev2024}, indicating the high quality and narrow linewidth of the material.

Based on these data we can now verify the theoretical assumption for the SHG. For this we calculate the dispersion at an angle of $21\degree$ from the magnetic field $f_{k_{21\degree}}(k)$ which is almost flat as shown in Fig.~\ref{fig:DISPERSIONS} (green line). By that logic, the line which satisfies energy and momentum conservation for SHG $2f_{k_{21\degree}}(k/2)$ is also flat but has twice the frequency (dashed line). A likely candidate to accommodate the higher harmonic magnons, is the first order perpendicular standing spin wave $f_{\text{PSSW}_{21\degree}}(k)$ which is plotted in blue. The blue and the dashed lines are almost on top of each other at lower frequencies so that for a range of $k$-vectors the condition for SHG is fulfilled. For comparison, the DEM dispersion is plotted in orange.

\subsection{Nonlinearity}\label{subsec3}
\begin{figure}
    \centering
    \includegraphics[width=1.0\linewidth]{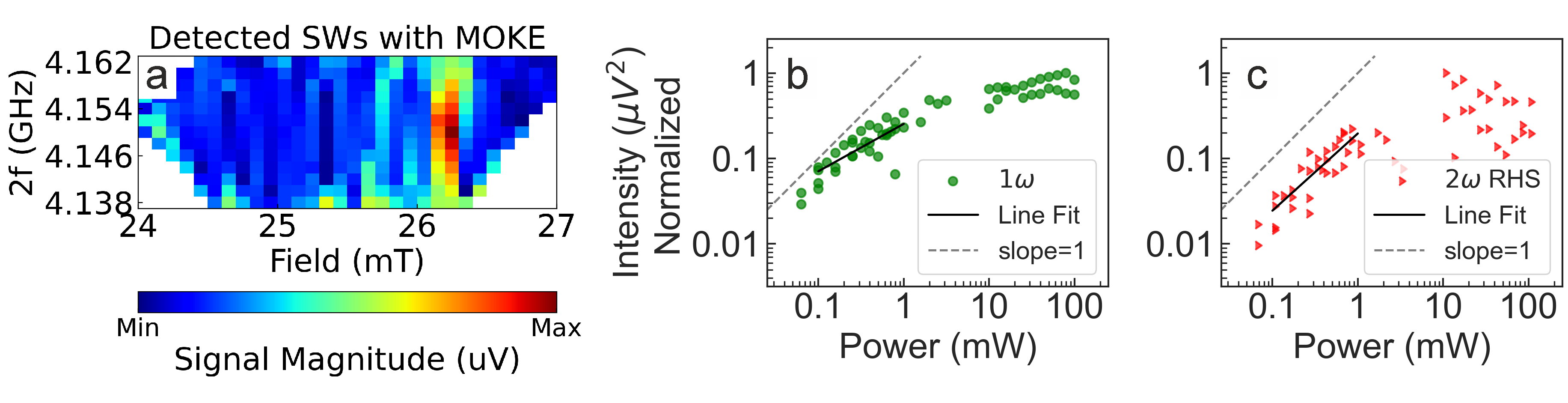}
    \caption{(a) Heatmap of 2f in the focus indicating the frequency and field range for which SHG is observed. (b) and (c) show in a double logarithmic plot how the detected intensity changes with excitation power for 1f and 2f, respectively. The black line serves as a guide for the eye that shows that the 2f intensity grows with a larger exponent than the 1f intensity. The dashed line indicates exponent 1. All intensities were averaged over 3x3 measurement points to reduce the noise. }
    \label{fig:SWEEPS}
\end{figure}

Plotting the power dependence of the measured intensities in the focus for $1\omega$ and $2\omega$ confirms the non-linear nature of the process. While for a linear process the intensities P$_{1f}$ and P$_{2f}$ should be strictly proportional, a non-linear process results in $P_{2f}\propto P_{1f}^n$ with $n>1$. In our measurements we observe $P_{2f}\propto P_{1f}^{1.5}$ which is super-linear and indicative of a non-linearity. Because the FFT shows a wide range of $k$-vectors for the second harmonic we do not expect the perfect quadratic behavior predicted by theory.
The fact that P$_{1f}$ is less than proportional to the excitation power, can be attributed to non-linear losses that happen before the concentrated spin waves reach the focus. It should be noted that at very large power, the losses start to dominate and both signals saturate while at very low power the noise becomes too large to allow for a precise analysis.

\section{Conclusions and Outlook}\label{sec5}
We have demonstrated the realization of a magnon concentrator design in lithographically patterned YIG structures that allows to locally create from a low-intensity Damon-Eshbach mode two concentrated high-intensity magnon beams with a new $k$-vector which overlap in a focal point. In this focal point the intensity can be more than 500 times higher than for the incoming magnons.  These large magnon intensities together with the new $k$-vector in the concentration area offer a platform for the localized generation of second harmonics. In principle, this approach is not limited to the creation of the second harmonic, but rather matching the dispersion with higher order modes may enable generation of even higher order harmonics. In particular, as the high intensity of the fundamental mode and the SHG are concentrated in an isolated spot even with low input RF-power levels, this indicates that a large variety of experiments requiring localized magnon power or SHG are now possible without need for a high intensity source.

\section{Methods}\label{sec:methods}
\subsection{Sample Fabrication}

The devices are fabricated using standard electron beam lithography processes and room-temperature pulsed laser deposition (PLD) of YIG films onto GGG (111) substrates. After the deposition and lift-off process, a designed structure of amorphous YIG remains, which crystallizes after an annealing process in oxygen as outlined in the literature \cite{Hauser2016}. After annealing and before the patterning of on-chip excitation waveguides, the samples undergo a wet-etch in phosphoric acid to reduce edge roughness resulting from the lift-off process.

\subsection{Simulations}
\subsubsection{Analytical spin wave dispersion}
When comparing the simulation to our experimental results  we use values $M_{sat} = 145000\,A/m$ and $t_{YIG} = 96.2\,nm$, and an exchange constant value $A = 3.5\,pJ/m$.\\
\subsubsection{Micromagnetic simulations}
The concentrator devices are modelled via $\mathsf{MuMax3}$ micromagnetic simulations. In the simulations, the concentrator structures are broken into cells of size $125\,nm$ x $125\,nm$ x $25\,nm$, where the film thickness and concentrator parameters $L_{fun}$, $\Theta_{fun}$, $d_{fun}$ can be set. The magnetic parameters are typical for YIG, specifically $M = 140\,kA/m$, $\alpha = 1\times 10^{-4}$. A constant magnetic field is applied and an excitation RF magnetic field is added based on an analytical solution of the Biot-Savart law for a stripline antenna set in a specific region. This enables us to reach higher $k$ values while still maintaining reasonable RF-field strength for given input power. The system is allowed to run for a sufficiently long enough time to achieve magnon propagation through the concentrator and establish steady state conditions throughout the system. The magnetization components are saved for various time steps and show the dynamic response of the devices as described in the main text.

\subsubsection{SNS-MOKE}
In scanning super-Nyquist sampling magneto-optical Kerr Effect microscopy (SNS-MOKE) we use ultrashort laser pulses to sample magnetization dynamics locally. In particular, the out-of-plane component of the dynamic precessional motion is detected. For this, we employ a frequency doubled Ti:Sa Laser operated at 80 MHz repetition rate. The non-linear frequency conversion to 400 nm allows for a better signal-to-noise ratio in YIG and offers a diffraction limited lateral spatial resolution of approximately 200 nm. We drive the magnetization dynamics in the YIG film with an RF-source synchronized with the laser repetition rate. This allows to perform lock-in demodulation at alias frequencies $f_{det} = |n \cdot f_{rep} - f_{RF}|$ due to undersampling of the GHz dynamics with a MHz sampling rate. In doing so, the response obtained by the magneto-optical Kerr effect (MOKE) is proportional to the dynamic susceptibility and yields amplitude and phase information of the precessional motion within the magnetic material. In addition, this technique is not limited by the detection of one frequency. The aliasing effects also result in additional frequency components e.g. for the second harmonic $f_{SHG} = |n \cdot f_{rep} - 2\cdot f_{RF}|$. Thus, SNS-MOKE can be used in combination with a multi-frequency lock-in amplifier to detect linear and non-linear dynamics simultaneously.

\backmatter
\bmhead{Supplementary information}
The supplementary material shows different scenarios of a spin wave approaching an interface under different respective angles. In addition it describes all parameters of the samples and measurements shown in the manuscript. 
\bmhead{Acknowledgements}
This work was partly funded by the Deutsche Forschungsgemeinschaft in project SCHM 2684/13-1,  by the German Research Foundation (DFG) as part of the German Excellence Strategy – EXC3112/1 – 533767171 (Center
for Chiral Electronics) and co-funded by the European Union under Grant Agreement No. 101070347 (MANNGA project).  
\bmhead{Author contributions}S.R. Lake did all the measurements, most of the data analysis and wrote part of the manuscript. M. Eger did further data analysis and simulations and prepared some of the figures, P. Geyer had the initial idea for the funnel structure, R. Dreyer did preliminary SNS-MOKE and worked on the manuscript, S.W. Kurfman worked on the interpretation and strongly contributed to the manuscript, G. Schmidt planned and supervised the experiments, contributed to the analysis and finalized the manuscript.  

\bmhead{Data availability statement}
Data openly available in a public repository that issues datasets with DOIs. The data that support the
findings of this study are openly available in Zenodo at http://dx.doi.org/10.5281/zenodo.20119412.

\bibliography{SW_Concentrators}

\end{document}


\title[Article Title]{SUPPLEMENTAL MATERIAL: Spatially-Localized Second Harmonic Generation via Spin Wave Concentration in Patterned YIG Structures}

\author[1]{\fnm{Stephanie R.} \sur{Lake}}

\author[1]{\fnm{Marc} \sur{Eger}}

\author[1]{\fnm{Philipp} \sur{Geyer}}

\author[1]{\fnm{Rouven} \sur{Dreyer}}

\author[1]{\fnm{Seth W.} \sur{Kurfman}}

\author*[1,2,3]{\fnm{Georg} \sur{Schmidt}}\email{georg.schmidt@physik.uni-halle.de}

\affil[1]{\orgdiv{Institut für Physik}, \orgname{Martin-Luther-Universität Halle-Wittenberg}, \orgaddress{\city{Halle}, \postcode{06120}, \country{Germany}}}
\affil[2]{\orgdiv{Interdisziplinäres Zentrum für Materialwissenschaften}, \orgname{Martin-Luther-Universität Halle-Wittenberg}, \orgaddress{\city{Halle}, \postcode{06120}, \country{Germany}}}
\affil[3]{\orgdiv{Halle-Berlin-Regensburg Cluster of Excellence CCE} \country{Germany}}


\maketitle
\begin{figure}
    \centering
    \includegraphics[width=1.0\linewidth]{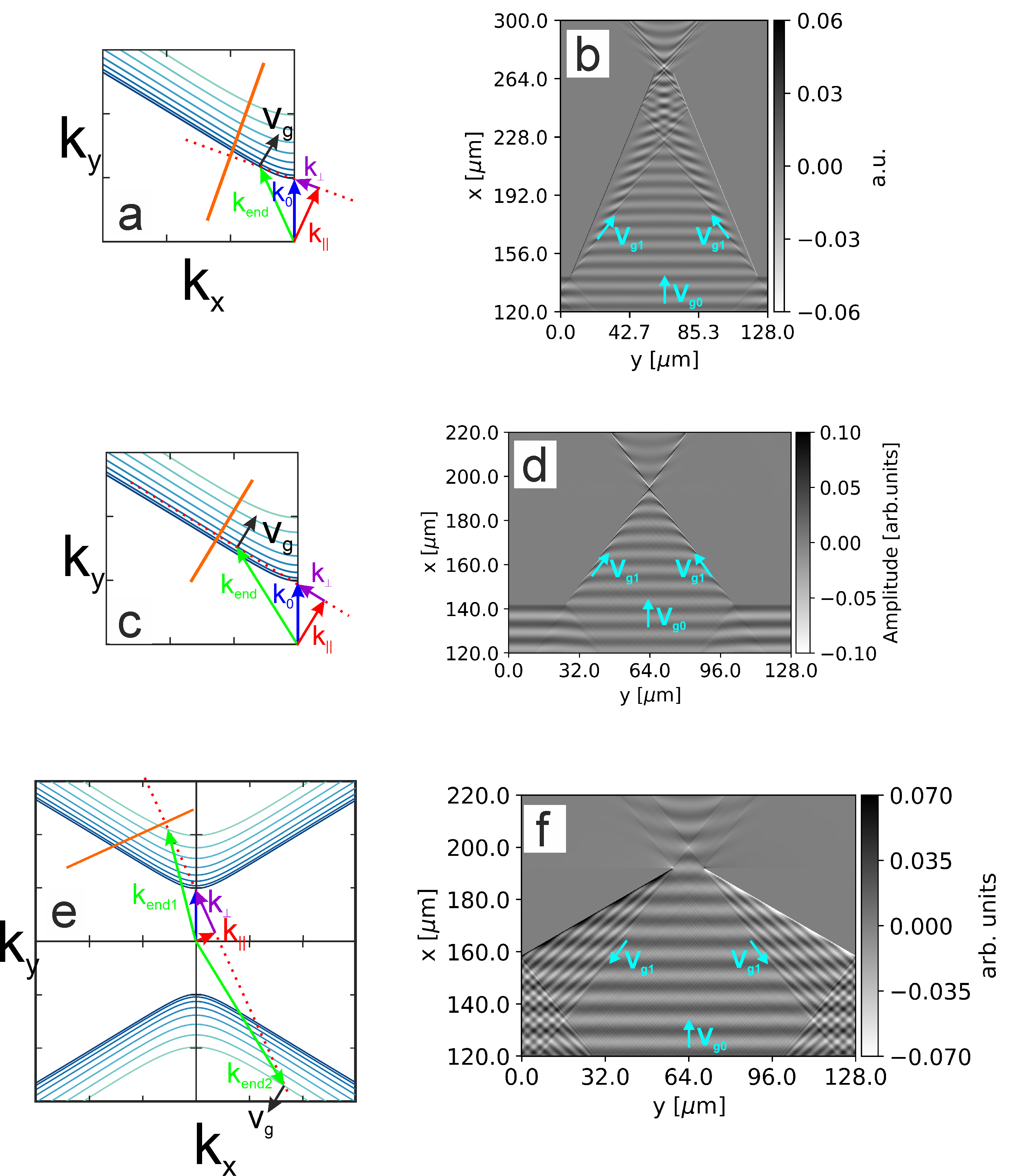}
    \caption{Left: schematic drawing of iso-frequency lines and change of $k$-vector and group velocity in k-space. Right: micromagnetic simulations f the magnon propagation in the respective funnel structures. $v_{g0}$ shows the initial group velocity, $v_{g1}$ indicates the group velocity after interaction with the boundary. (a) If the angle $\Theta$ is too small, the dotted line can only be tangent to one iso-frequency line in one single point and then leads back towards iso-frequency lines for larger magnetic fields. This is completely consistent with $\vec{v}_g$ turning first towards parallel to the interface but than turning away from it again. So the magnons are not collected on a single line but simply reflected inward as is also shown in the simulation (b). (c) and (d) show the correct conditions for focusing.
    For $\Theta$ being too large (e), the situation is more complex. The dotted line can never be tangent to an iso-frequency line and at some point the boundary is reached. $k_{||}$ is still conserved while $k_{\perp}$ is reversed, but still has to be on an iso frequency line for the same magnetic field. $\vec{k}$ then lies in the opposite quadrant but is not point-symmetric with respect to the origin. The resulting wave has a k-vector pointing backward and towards the optical axis, while $\vec{v}_g$ also points backward but away from the optical axis as shown in the simulation (f).
 }
    \label{scenarios}
\end{figure}

Besides the optimum operating conditions, it is also worth noting what happens when the angle $\Theta$ is either too small or too large. Different cases are shown in fig. S\ref{scenarios}. 

An important point for measurements and simulations is how the magnitude and its square the intensity are determined.
For the experiment, the magnitude is given by $R = \sqrt{X^2+Y^2}$  where $X$ and $Y$ are the corresponding channels of the lock-in demodulators. For the simulation the magnitude is $M= \sqrt{m_x^2+ \alpha_y m_y^2}$ with $\alpha_y$ being a pre-factor to appropriately minimize artifacts in the plotting  related to the spin wave precession ellipticity. 

The experimental parameters for Fig. 4 (sample A) were $f$ = 3.51 GHz, $B$ = 62.81 mT, $d_{gap} = 12 \mu m$, $\Theta_\text{fun} = 35\degree$, and an applied power to the excitation antenna $P$ = 1 mW. For the simulations, the parameters were $f$ = 3.51 GHz, $B$ = 62.81 mT, $d_{gap} = 12 \mu m$,  $\Theta_\text{fun} = 35\degree$, and an effective excitation power of $P$ = 1.0125 mW was applied.
For the determination of the focusing factors the values at (experiment: 2.5, -50.2) and (simulation: 0.0, -54.0) were used to normalize, respectively.

For Fig. 5 (a) (sample B), the experimental parameters are $f$ = 3.685 GHz, $B$ = 63.76 mT, and an applied power to the excitation antenna $P$ = 1 mW. The device parameters are $d_{gap} = 6 \mu m$ and $\Theta_{fun} = 33\degree$.
For the simulations $f$=3.25 GHz, $B$ = 52.62 mT, $d_{gap} = 6µm$, $\theta_{fun}= 35\degree$ and $P$ = 1.0125 mW were used.

For the linear magnons shown in Fig.6, the experimental parameters are $f$ = 2.080 GHz, $B$ = 26.37 mT, and an applied power to the excitation antenna $P$ = -3 dBm. The device parameters are $d_{gap} = 10 \mu m$ and $\Theta_{fun} = 35\degree$. For the second harmonics shown in subfigs. c-f, all parameters are the same except for the SNS-MOKE detecting frequency, as that has been set to twice the frequency of the linear magnons, 4.160 GHz.